\documentclass[prl,letterpaper,twocolumn,noshowpacs,floatfix,groupedaddress,amsmath,amsfonts,amssymb,preprintnumbers,nofootinbib,citeautoscript]{revtex4}
\pdfoutput=1
\usepackage[T1]{fontenc}\usepackage[latin1]{inputenc}
\usepackage{dcolumn,graphicx,color,hvfloat,booktabs,wrapfig,microtype,afterpage} \graphicspath{{Figures/}}
\usepackage[charter,greekuppercase=italicized]{mathdesign}
\usepackage{sidecap}
\renewcommand{\figurename}{\bf\textsf{Figure}\sffamily}
\renewcommand{\tablename}{Table}
\makeatletter\renewcommand{\fnum@figure}[1]{\textbf{\sffamily\figurename~\thefigure~|\,}}\makeatother
\makeatletter\renewcommand{\fnum@table}[1]{\textbf{\sffamily\tablename~\thetable~|\,}}\makeatother

\newcommand{\half}{\frac{1}{\protect\raisebox{0.7pt}{$\scriptstyle 2$}}}
\newcommand{\quarter}{\frac{1}{\protect\raisebox{0.7pt}{$\scriptstyle 4$}}}

\newcount\hh \newcount\mm
\hh=\time \divide\hh by 60
\mm=\hh \multiply\mm by 60 \mm=-\mm
\advance\mm by \time
\def\now{\number\hh:\ifnum\mm<10{}0\fi\number\mm}

\definecolor{NatureBlue}{rgb}{0.012,0.3,0.63}
\usepackage[colorlinks,plainpages=false,linkcolor=NatureBlue,urlcolor=NatureBlue,citecolor=NatureBlue,pdfpagemode=UseNone,pdfstartview=FitBH]{hyperref}

\makeatletter
\newcommand{\bibstyle@supplement}{\bibpunct[, ]{[S}{]}{;}{n}{,}{,}%
    \gdef\bibnumfmt##1{[S##1]}}
\makeatother

\begin{document}


\title{\flushleft\fontsize{22pt}{26pt}\selectfont\sffamily\bfseries\textcolor{NatureBlue}{Resonant magnetic exciton mode in the heavy-fermion antiferromagnet CeB$_6$}}

\author{\large\sffamily G.~Friemel$^1$} \author{Yuan~Li$^1$} \author{A.~V.~Dukhnenko$^2$} \author{N.~Yu.~Shitsevalova$^2$} \author{N.~E.~Sluchanko$^3$} \author{A.~Ivanov$^4$} \author{$\phantom{...\kern1.85pt}$V.~B.~Filipov$^2$} \author{B.\,Keimer$^1$} \author{D.\,S.\,Inosov$^{1,\,\ast^{\mathstrut}}$\hfill\smallskip}
\affiliation{\flushleft\mbox{\sffamily $^1$\hspace{0.5pt}Max-Planck-Institut für Festkörperforschung, Heisenbergstraße 1, 70569 Stuttgart, Germany.}
\mbox{\sffamily $^2$\hspace{0.5pt}I.\,M.\,Frantsevich Institute for Problems of Material Sciences of NAS, 3 Krzhyzhanovsky str., 03680 Kiev, Ukraine.}
\mbox{\sffamily $^3$\hspace{0.5pt}A.\,M.\,Prokhorov General Physics Institute of the Russian Academy of Sciences, 38 Vavilova str., 119991 Moscow, Russia.}
\mbox{\sffamily $^4$\hspace{0.5pt}Institut Laue-Langevin, 6 rue Jules Horowitz, BP 156, 38042 Grenoble Cedex, France.}}

\begin{abstract}\citestyle{nature}
\parfillskip=0pt\relax\fontsize{8.9pt}{11pt}\selectfont
\noindent\textbf{Resonant magnetic excitations are widely recognized as hallmarks of unconventional superconductivity in copper oxides,\cite{Eschrig06, HinkovBourges07} iron pnictides,\cite{InosovPark10} and heavy-fermion compounds.\cite{StockBroholm08, SatoAso01, StockertArndt10} Numerous model calculations have related these modes to the microscopic properties of the pair wave function,\cite{Chubukov08, Eremin08, OnariKontani10, MaierGraser09} but the mechanisms underlying their formation are still debated.\cite{Eschrig06} Here we report the discovery of a similar resonant mode in the non-superconducting, antiferromagnetically ordered heavy-fermion metal CeB$_6$.\cite{Effantin85, GoodrichYoung04, HallFisk00, NakaoMagishi01, TanakaStaub04, MatsumuraYonemura09} Unlike conventional magnons, the mode is non-dispersive, and its intensity is sharply concentrated around a wave vector separate from those characterizing the antiferromagnetic order. The magnetic intensity distribution rather suggests that the mode is associated with a coexisting order parameter of the unusual antiferro-quadrupolar phase of CeB$_6$, \cite{NakaoMagishi01, TanakaStaub04, MatsumuraYonemura09} which has long remained ``hidden'' to the neutron-scattering probes. \cite{Effantin85, GoodrichYoung04} The mode energy increases continuously below the onset temperature for antiferromagnetism, \textit{T}$\!_{\bf N}$, in parallel to the opening of a nearly isotropic spin gap throughout the Brillouin zone. These attributes bear strong similarity to those of the resonant modes observed in unconventional superconductors\cite{Eschrig06, HinkovBourges07, StockBroholm08, SatoAso01, StockertArndt10, InosovPark10} below their critical temperatures, \textit{T}$\!_{\bf c}$. This unexpected commonality between the two disparate ground states indicates the dominance of itinerant spin dynamics in the ordered low-temperature phases of CeB$_{\bf 6}$ and throws new light on the interplay between antiferromagnetism, superconductivity, and ``hidden'' order parameters in correlated-electron materials.\vspace{-3pt}}
\end{abstract}

\pacs{78.70.Nx 75.30.Mb 75.30.-m 75.25.Dk\vspace{-0.7em}}

\pagestyle{plain}
\makeatletter
\renewcommand{\@oddfoot}{\hfill\bf\scriptsize\textsf{\thepage}}
\renewcommand{\@evenfoot}{\bf\scriptsize\textsf{\thepage}\hfill}
\makeatother

\citestyle{nature}
\maketitle\enlargethispage{0.5em}

\makeatletter\immediate\write\@auxout{\string\bibstyle{my-nature}}\makeatother
\renewcommand\bibsection{\section*{\sffamily\bfseries\footnotesize References\vspace{-10pt}\hfill~}}

\noindent
In copper-oxide, iron-pnictide, and heavy-fermion superconductors, antiferromagnetic (AFM) order can be induced by changes of pressure, doping level, or magnetic field. Theories of magnetically mediated Cooper pairing inspired by this observation have received strong support from neutron scattering experiments that revealed ``resonant'' magnetic excitations in the superconducting states of all of these compounds.\cite{Eschrig06, HinkovBourges07, StockBroholm08, SatoAso01, StockertArndt10, InosovPark10} The resonant modes share a common set of characteristics. They are sharp in both energy and momentum, unlike the incoherent particle-hole excitations in a metal, and are located below the superconducting gap, $2\Delta$. Since model calculations indicate that the spectral weight and momentum-space structure of the resonant modes contain essential information about the pairing interaction,\cite{Eschrig06, StockertArndt10, InosovPark10} pairing symmetry,\cite{Chubukov08, Eremin08, OnariKontani10, MaierGraser09} and condensation energy\cite{WooDai06} of the superconducting state, research on their origin remains at the frontier of the quest for a comprehensive description of unconventional superconductivity.

\begin{figure}[b]\vspace{-0.1em}
\includegraphics[width=\columnwidth]{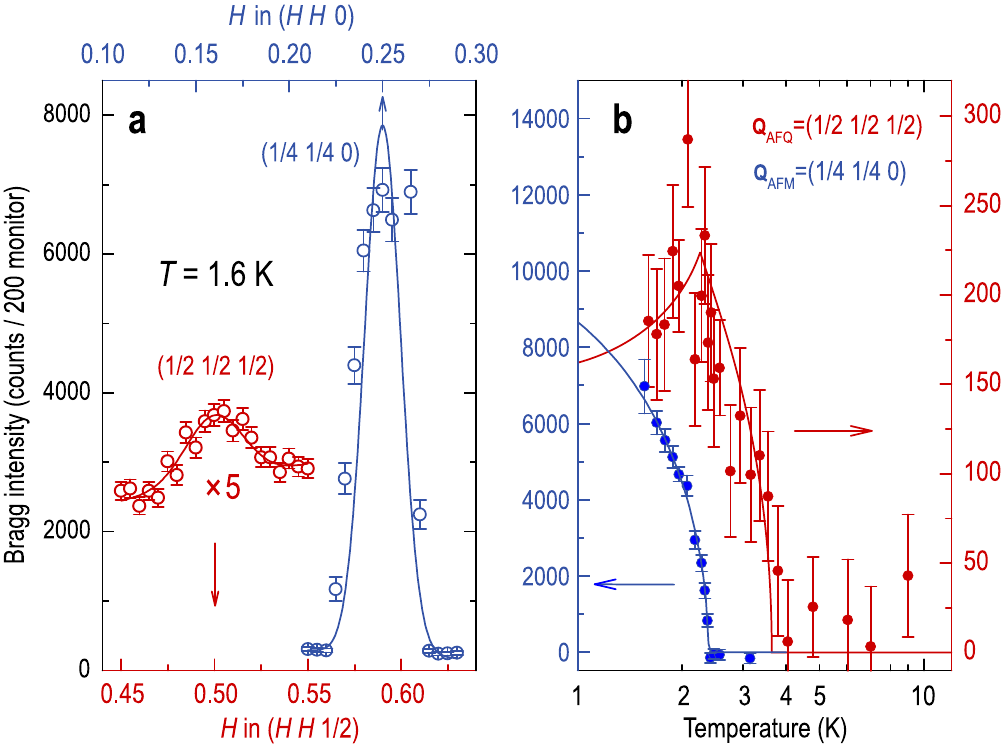}
\caption{\textbf{Bragg peaks associated with the AFQ and AFM order parameters.} \textbf{a}, Comparison of the elastic signals measured at $\mathbf{Q}_{\rm AFQ}=R(\half \half \half)$ and $\mathbf{Q}_{\rm AFM}=S(\quarter \quarter 0)$ wave vectors at $T=1.6$\,K. \textbf{b}, Temperature dependence of the peak intensities that vanish at $T_{\rm Q}$ and $T_{\rm N}$, respectively. The solid lines are guides to the eyes.\vspace{-1.5em}}
\label{Fig:Elastic}
\end{figure}

\begin{figure*}[t]\vspace{-4pt}
\includegraphics[width=\textwidth]{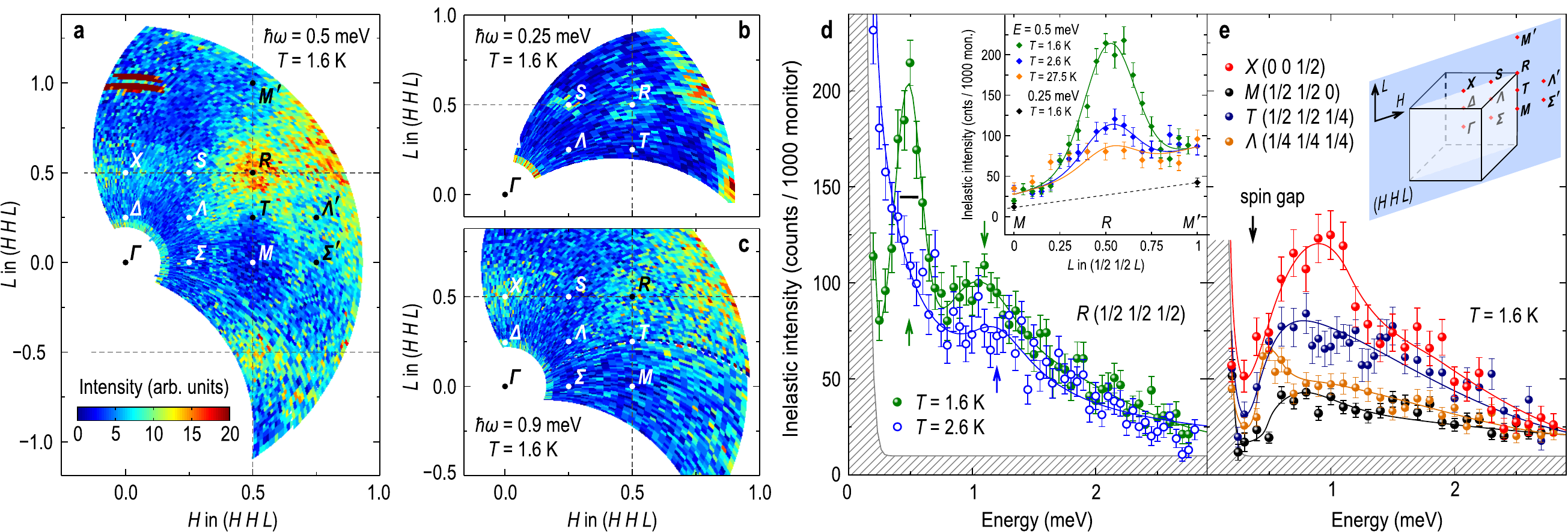}
\caption{\textbf{Unprocessed INS data.} \textbf{a}, Inelastic signal along the $(HHL)$ plane in reciprocal space, measured in the low-temperature AFM phase ($T=1.6$\,K) at 0.5\,meV energy transfer. The labels represent different high-symmetry points, and the dashed lines mark Brillouin zone boundaries. \textbf{b,\,c}, The same at lower (0.25\,meV) and higher (0.9\,meV) energies, respectively. \textbf{d}, Energy dependence of the INS signal in the AFQ (empty symbols) and AFM (filled symbols) states, measured at the $R(\half \half \half)$ point, where the intensity is maximized. Exciton peaks are marked by arrows. The horizontal black bar inside the peak shows the experimental energy resolution, the hatched region represents background intensity, and the lines are guides to the eyes. The inset shows constant-energy scans across the peak maximum ($\hslash\omega=0.5$\,meV) along the $MRM^\prime$ line at different temperatures; the dashed line is the background level suggested by the intensity minima at $\smash{M(\half \half 0)}$ and $\smash{M^\prime(\half \half 1)}$ within the spin gap ($\hslash\omega=0.25$\,meV). \textbf{e}, Low-temperature ($T=1.6$\,K) energy scans for the $X$, $M$, $T$, and $\Lambda$ points. The location of the scattering plane and the naming of high-symmetry points in the cubic Brillouin zone are shown in the inset.\vspace{-1.8em}}
\label{Fig:RawData}
\end{figure*}

Here we report inelastic neutron scattering (INS) experiments that reveal a resonant mode with identical characteristics in the non-superconducting heavy-fermion metal CeB$_6$, which has been intensely investigated because of its intriguing phase behavior\cite{Effantin85, GoodrichYoung04, HallFisk00} on the background of an exceptionally simple crystal structure and chemical composition. The electronic states near the Fermi level of CeB$_6$ are composed of localized cerium-4$f^1$ levels hybridized with itinerant cerium-5$d$ and boron-2$p$ electrons,\cite{GrushkoPaderno85, PlakhtyRegnault05} closely analogous to those of the superconductors CeCoIn$_5$ and CeCu$_2$Si$_2$, where prominent resonant modes have been reported.\cite{StockBroholm08, StockertArndt10} In contrast to these superconducting compounds, however, CeB$_6$ exhibits AFM order below $T_{\rm N}=2.3$\,K.\cite{ZaharkoFischer03} The AFM transition is preceded by another phase transition at $T_{\rm Q}=3.2$\,K, whose order parameter has long remained ``hidden'' to standard experimental probes such as neutron diffraction. \cite{Effantin85, GoodrichYoung04} Theoretical work \cite{ShiinaShiba97, ThalmeierShiina03} has attributed both transitions to interactions between the multipolar moments of the Ce-$4f$ electrons mediated by the itinerant conduction electrons, which break the large ground-state degeneracy of the cerium ions in their cubic crystal field and stabilize an antiferro-quadrupolar (AFQ) order. \cite{NakaoMagishi01, TanakaStaub04, MatsumuraYonemura09}

Recent transport, thermodynamic, and neutron diffraction experiments \cite{SluchankoBogach07, SluchankoBogach08, DemishevSemeno09, AnisimovBogach09, PlakhtyRegnault05} have revealed deviations from predictions of this ``interacting multipole'' model of magnetism in CeB$_6$. In particular, weak Bragg intensity has been detected by spin-polarized neutron scattering at the AFQ wave vector, although the concept of quadrupolar ordering forbids such an intensity in the absence of external magnetic fields, \cite{Effantin85, GoodrichYoung04} so that it is ``hidden'' to the elastic neutron scattering. This observation has been ascribed to the magnetic response of the itinerant Ce-$5d$ electrons.\cite{PlakhtyRegnault05} Our discovery of a magnetic resonant mode for $T <T_{\rm N}$ confirms a more pronounced influence of itinerant electrons on the magnetic properties of CeB$_6$ than anticipated by the prevailing theory, and establishes a surprising correspondence to unconventional superconductivity.

We start the presentation of our data by comparing the elastic neutron scattering intensities measured at the previously identified AFM and AFQ Bragg peaks at $\mathbf{Q}_{\rm AFM}=S(\quarter \quarter 0)$ and at $\mathbf{Q}_{\rm AFQ}=R(\half \half \half)$, respectively, in Fig.\,\ref{Fig:Elastic}. From the different onset temperatures of the elastic signals, it is clear that they represent distinct order parameters. The elastic intensity at $\mathbf{Q}_{\rm AFQ}$ is more than an order of magnitude weaker than the AFM Bragg intensity, and exhibits only a weak suppression below $T_{\rm N}$, indicating that both order parameters coexist in the AFM phase.

To give an overview of the inelastic-scattering signal, we present in Fig.\,\ref{Fig:RawData}a--c unprocessed INS intensity maps of the $(HHL)$ plane in reciprocal space, measured in the AFM low-temperature phase ($T=1.6$\,K) at fixed energy transfers of $\hslash\omega=0.25$, 0.5, and 0.9 meV. An intense peak centered at $\mathbf{Q}_{\rm AFQ}$ dominates the spectrum at $\hslash\omega=0.5$ meV. It can be observed at the two equivalent $R$ points $(\half \half \half)$ and $(\half \half -\!\!\half)$ covered by the measurement, which are additionally connected by weaker streaks of intensity that surround the $M$ point. Intensity maps at both lower (Fig.\,\ref{Fig:RawData}b) and higher (Fig.\,\ref{Fig:RawData}c) energies reveal greatly reduced intensities. The strong concentration of magnetic spectral weight around isolated points in energy-momentum space is markedly different from the continuously dispersive spin waves in conventional antiferromagnets, but closely similar to the behavior of the resonant modes observed in copper-oxide, iron-pnictide, and heavy-fermion superconductors below their respective transition temperatures. \cite{Eschrig06, HinkovBourges07, SatoAso01, StockBroholm08, StockertArndt10, InosovPark10} Moreover, the mode is not centered at the $S$ or $\Sigma$ points characterizing the AFM order, where the magnetic intensity is much weaker, but at the $R$ point that reflects the ``hidden'' AFQ phase.

\begin{figure*}[t]
\includegraphics[width=\textwidth]{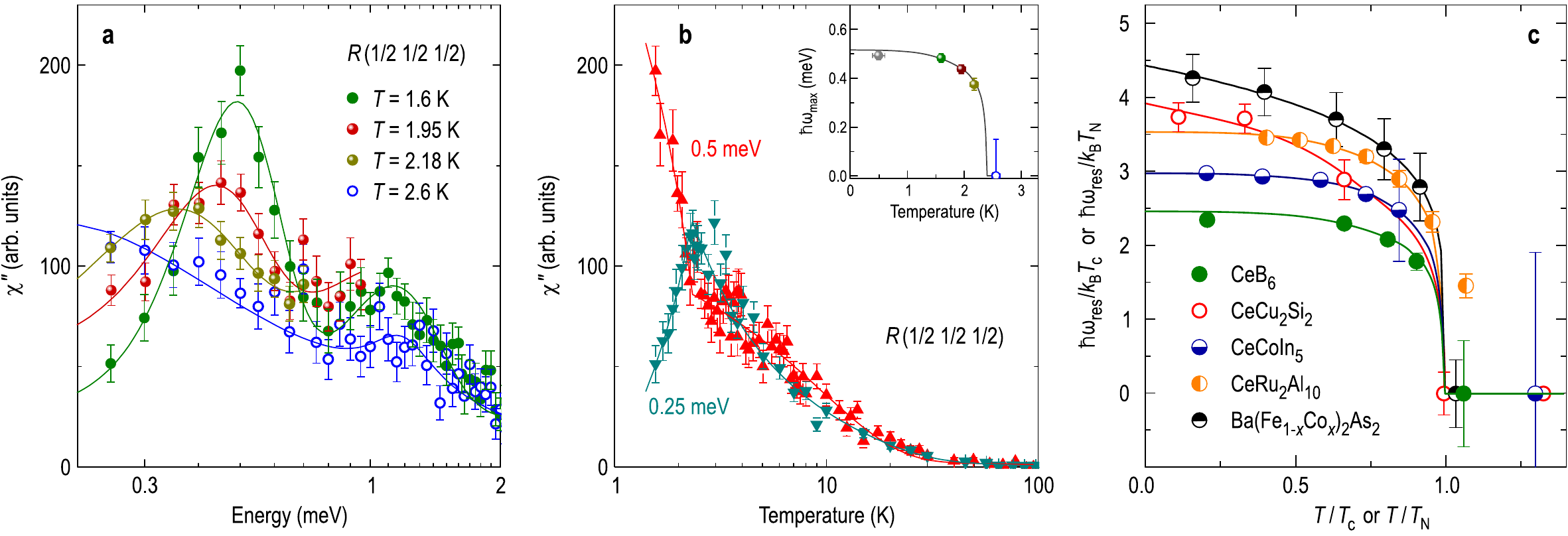}
\caption{\textbf{Temperature dependence.} \textbf{a}, Imaginary part of the dynamic spin susceptibility, $\chi''(\mathbf{Q}_{\rm AFQ},\,\omega)$, at various temperatures, illustrating the gradual development of the spin-exciton peak below the AFM transition. The lines are visual guides. \textbf{b}, Temperature dependence of $\chi''(\mathbf{Q}_{\rm AFQ},\,\omega)$ at the peak maximum ($\hslash\omega=0.5$\,meV) and within the spin gap ($\hslash\omega=0.25$\,meV). The inset shows the exciton energy versus temperature, determined from the position of the peak maximum, $\hslash\omega_{\rm max}$, in panel a. \textbf{c}, Comparison of the normalized resonance energies, $\hbar\omega_{\rm res}/k_{\rm B}T_{\rm N}$ (or $T_{\rm c}$), plotted vs reduced temperature, $T/T_{\rm N}$ (or $T_{\rm c}$), in CeB$_6$, in several unconventional superconductors,\cite{StockertArndt10, StockBroholm08, InosovPark10} and in the low-temperature phase of CeRu$_2$Al$_{10}$.\cite{RobertMignot10} The lines are guides to the eyes.\vspace{-1.8em}}
\label{Fig:Tdep}
\end{figure*}

At $T=1.6$\,K, the resonant mode is sharp in energy (Fig.\,\ref{Fig:RawData}d) and is characterized by the intrinsic (resolution-corrected) full width at half maximum (FWHM) of $\sim$\,0.2\,meV. It is followed by a weaker peak centered around 1.1\,meV, with the FWHM of $\sim$\,0.6\,meV. For $T > T_{\rm N}$, the 0.5\,meV peak transforms into a quasielastic response \cite{HornSteglich81, BouvetThesis} that is typical for the paramagnetic state of heavy-fermion metals.\cite{SatoAso01, StockertArndt10} We can therefore associate the formation of the resonant peak with the opening of a spin gap in the low-energy part of the quasielastic spectrum upon cooling and the consequent spectral-weight transfer from below $\sim$\,0.35\,meV to higher energies. In contrast, the broader peak at 1.1\,meV is nearly insensitive to $T_{\rm N}$ and is possibly due to a low-energy crystal-field excitation similar to those previously studied in the AFQ phase. \cite{RegnaultErkelens88, BouvetThesis}

The detailed temperature dependence of the resonant mode is illustrated by Fig.\,\ref{Fig:Tdep}, where we plot the imaginary part of the dynamical spin susceptibility, $\chi''(\mathbf{Q}_{\rm AFQ},\omega)$, obtained by correcting the background-subtracted INS signal for the thermal population factor. In Fig.\,\ref{Fig:Tdep}a, one sees that the mode gradually broadens and shifts towards lower energies upon warming, following the order-parameter-like trend shown in the inset of Fig.\,\ref{Fig:Tdep}b. This behavior is again strikingly similar to the one of the resonant mode in unconventional superconductors (Fig.\,\ref{Fig:Tdep}c). For $T > T_{\rm N}$, the quasielastic signal continues to diminish in intensity with increasing temperature (Fig.\,\ref{Fig:Tdep}b). Constant-energy scans in the inset of Fig.\,\ref{Fig:RawData}d demonstrate that it always remains peaked at the commensurate wave vector $\mathbf{Q}_{\rm AFQ}$, where the resonant mode develops below $T_{\rm N}$.

Constant-$\mathbf{Q}$ scans at various high-symmetry points (Fig.\,\ref{Fig:RawData}e) further show a weakly energy dependent magnetic response throughout the Brillouin zone. For $T < T_{\rm N}$, a spin gap of magnitude $\sim 0.5$\,meV opens up nearly isotropically in momentum space, including the $M$ point where the magnetic INS signal is minimal, and the $R$ point where the resonant mode develops in the AFM state. The appearance of the spin gap is consistent with the depletion in the electronic density of states below $T_{\rm N}$, previously observed by point-contact spectroscopy. \cite{Kunii87} The comparison in Fig.\,\ref{Fig:Tdep}b of the temperature scans measured at the peak maximum (0.5\,meV) and below it (0.25\,meV) emphasizes that the spectral-weight gain at the exciton energy is compensated by its depletion within the spin-gap region. Both effects show a simultaneous sharp onset at $T_{\rm N}$, closely similar to the behavior of the resonant mode and the spin gap in an unconventional superconductor.\cite{InosovPark10} An analogous $T$-evolution is observed at the $X(0 0 \half)$ point (see Fig.\,\ref{Fig:S2} in the Supplementary Information), where a quasielastic signal of comparable intensity is found in the AFQ state. The redistribution of spectral weight at this point leads to an intensity enhancement below $T_{\rm N}$ at $\hbar\omega\approx0.9$\,meV, but the resulting feature is much broader than the one at the $R$ point.

\begin{figure}[t]\vspace{-0.1em}
\includegraphics[width=0.8\columnwidth]{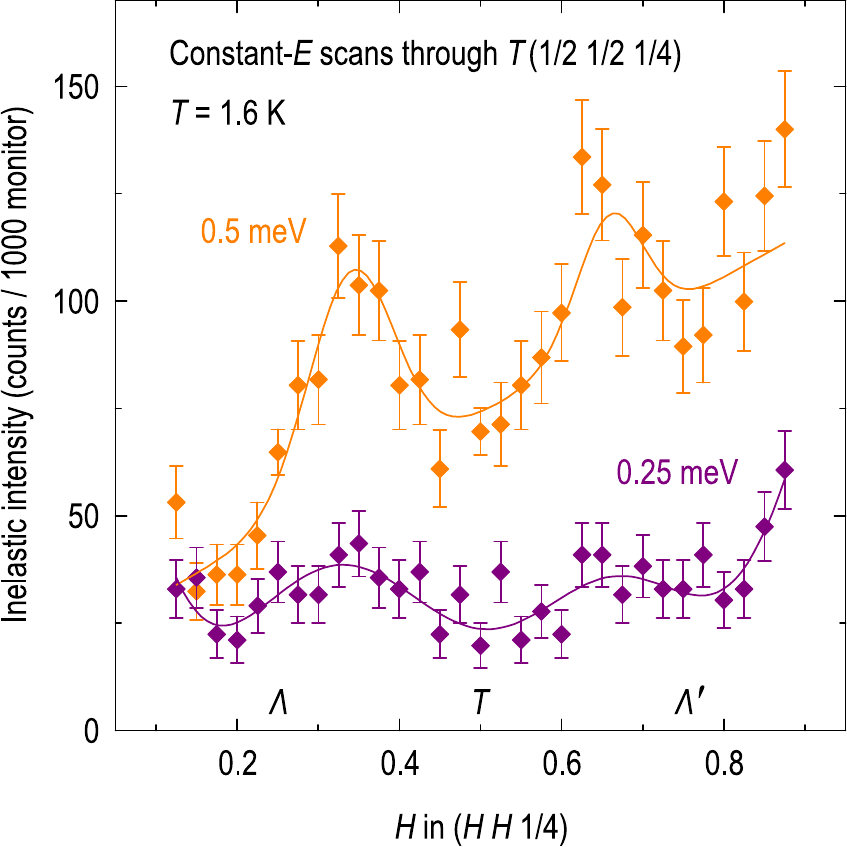}
\caption{\textbf{Anisotropic shape of the resonant mode in momentum space.} Low-temperature $\mathbf{Q}$-scans along the $\Lambda T\!\Lambda^\prime$ direction cross the two lobes of the resonant intensity, resulting in two incommensurate peaks. The intensity is maximized at the resonance energy ($\hslash\omega=0.5$\,meV) and is much reduced within the spin gap ($\hslash\omega=0.25$\,meV).\vspace{-1.8em}}
\label{Fig:QScans}
\end{figure}

Although the resonant mode is most intense at $R(\half \half \half)$, Fig.\,\ref{Fig:RawData}a suggests the presence of weaker tails extending well into the Brillouin zone. This observation is substantiated by Fig.\,\ref{Fig:QScans}, where we show momentum scans along the $\Lambda T\!\Lambda^\prime$ direction that cut the two lobes of intensity extending from $R$ towards the $M$ point (see also Fig.\,\ref{Fig:S2}b in the Supplementary Information). The comparison of the low-temperature signals measured at 0.5 and 0.25\,meV together with the energy scan at the $\Lambda$ point (Fig.\,\ref{Fig:RawData}e) let us conclude that the mode exhibits at most a weak dispersion, so that even away from the $R$ point its intensity remains maximal near $\hslash\omega_{\rm max}\approx0.5$\,meV.

While the quasielastic intensity in the paramagnetic state of CeB$_6$ is similar to the one observed in other heavy-fermion compounds,\cite{SatoAso01, StockertArndt10} the behavior of the resonant mode we have discovered for $T < T_{\rm N}$ is quite different from that of low-energy magnons in other heavy-fermion antiferromagnets including the isostructural PrB$_6$ \cite{LeMcEwen08} as well as CeIn$_3$ \cite{CeIn3} and CePd$_2$Si$_2$. \cite{CePd2Si2} These excitations exhibit the conventional magnon dispersion with minimum energy and maximum intensity at the AFM ordering wave vector. The resonant mode in CeB$_6$, on the other hand, is much more intense, at most weakly dispersive, and centered at $\mathbf{Q}_{\rm AFQ}$ rather than $\mathbf{Q}_{\rm AFM}$. The behavior we have observed disagrees with calculations in the framework of the interacting-multipole model of CeB$_6$, which predict magnon-like low-energy modes,\cite{ThalmeierShiina03} in addition to more weakly dispersive higher-energy crystal-field excitations.\cite{LoewenhauptCarpenter85}

We have already pointed out the striking similarity of the resonant mode in CeB$_6$ to the low-energy spin excitations in unconventional superconductors. While this analogy has not been anticipated as far as we are aware, we can understand it qualitatively in the framework of a two-component model that includes strong interactions of the localized magnetic moments with the itinerant electrons.\cite{SatoAso01} The formation of the AFM state accounts for the opening of a gap in the electronic density of states near the Fermi energy, which is similar to that found in spin-density-wave (SDW) systems. As a consequence, it leads to the nearly isotropic spin gap in the continuous magnetic spectrum originating from strongly hybridized crystal-field excitations of the Ce $4f$-electrons and the itinerant conduction electrons. The resonant mode can then be regarded as a collective mode below the onset of the particle-hole continuum, in close analogy to current theoretical models of the resonant modes in the superconductors. We note that this scenario is quite different from the standard interacting-multipole model of CeB$_6$, according to which the magnetic response in the AFM phase is dominated by the magnetic dipole moments of the localized Ce electrons, but qualitatively consistent with other recent observations that emphasized the importance of itinerant magnetism in this system. \cite{PlakhtyRegnault05, SluchankoBogach07, SluchankoBogach08, DemishevSemeno09, AnisimovBogach09}

While there is general agreement on the collective-mode nature of the resonant modes in unconventional superconductors, the mechanisms leading to their formation are still controversial.\cite{Eremin08, Chubukov08} It is interesting to consider our data on CeB$_6$ in the light of two classes of theories that describe the mode as a magnon-like excitation of localized electrons close to a magnetically ordered state, and as an excitonic bound state of itinerant electrons, respectively. A recent calculation in the framework of the former approach \cite{Chubukov08} has reproduced the temperature evolution of the magnetic response of superconducting CeCoIn$_5$ from an overdamped mode above $T_{\rm c}$ to a resonant collective mode below $T_{\rm c}$ by assuming that the superconducting gap removes damping channels for a low-energy Ce crystal-field excitation. An analogous model in which the superconducting gap is replaced by a SDW-type gap may explain the resonant mode in CeB$_6$. On the one hand, the appearance of the resonant mode at the $R$ point is consistent with the presence of AFQ order, which is expected to support a soft multipolar mode at this wave vector. On the other hand, the close proximity of the resonant mode to the gap energy, which would have to be regarded as a coincidence in this approach, is naturally explained in the framework of a scenario in which the mode is described as a weakly bound triplet exciton. The complex anisotropic shape of the collective mode in CeB$_6$ also suggests an influence of the fermiology of the conduction-electron system, in analogy to itinerant-electron descriptions of the resonant mode in CeCoIn$_5$ \cite{Eremin08} and the ``hourglass'' shape of the resonant mode in the copper oxides.\cite{Eschrig06} While such an influence is generally expected for an exciton within an itinerant-electron description, it will be interesting to explore theoretically the impact of the coherence factors on the expression for $\chi''(\mathbf{Q},\omega)$ in the AFM ordered state. This is important because the relationship between the superconducting coherence factors and the resonant mode has figured prominently in the ongoing debate about the Cooper pairing symmetry in iron pnictides \cite{OnariKontani10, MaierGraser09, InosovPark10} and heavy-fermion compounds \cite{Chubukov08, Eremin08}. Answers to these questions will have to await calculations that take full account of the strong coupling between the localized Ce crystal-field states and the itinerant electrons, qualitatively beyond the current interacting-multipole model of CeB$_6$.

We end our discussion by noting that the resonant mode in CeB$_6$ apparently results from the coexistence of the AFM state with the unusual AFQ order that had remained ``hidden'' to experimental probes in zero magnetic field until the recent advent of resonant x-ray diffraction methods.\cite{NakaoMagishi01, TanakaStaub04, MatsumuraYonemura09} This unique situation explains its unusual behavior in comparison to other heavy-fermion antiferromagnets. Presumably as a consequence of AFQ-order fluctuations, the dynamical susceptibility at the $R$-point of CeB$_6$ remains substantial even for $T > T_{\rm N}$, where the resonant mode is overdamped, and only decreases gradually upon further heating. This behavior is reminiscent of the gradual onset of magnetic intensity in the ``pseudogap'' state of the underdoped cuprates, \cite{HinkovBourges07} for which various ``hidden'' order parameters have also been proposed. The mode we have observed may also be related to the higher-energy excitonic modes that have been reported in other materials with thus-far unidentified ``hidden'' order parameters, such as CeRu$_2$Al$_{10}$, \cite{RobertMignot10} and in small-gap ``Kondo insulators'' SmB$_6$ \cite{AlekseevMignot95} and YbB$_{12}$, \cite{MignotAlekseev05} whose ground states remain poorly understood. Since the ``hidden'' order in CeB$_6$ has actually been extensively characterized, the comprehensive description of the magnetic dynamics of this compound we have initiated here will have model character for a broad class of correlated-electron materials.

\vspace{0.8em}
{\footnotesize\noindent
{\sffamily\textbf{Methods.}}\vspace{1ex}

\noindent Two single-crystalline rods of CeB$_6$ with a total mass of 8\,g were grown by the floating-zone method from a 99.6\,at.\,\% isotope-enriched $^{11}$B powder, in order to minimize the absorption of neutrons by the $^{10}$B isotope.

The data were collected using the cold triple-axis IN14 spectrometer (ILL, Grenoble, France) with a pyrolytic graphite (PG) monochromator. The sample was mounted into a standard cryostat with the (110) and (001) directions in the scattering plane. To acquire the overview maps of the reciprocal space (Fig.\,2a--c), we used the \textit{Flatcone} detector, equipped with Si\,(111) analyzers that select scattered neutrons with the finite momentum $k_{\rm f}=1.40$\,\AA$^{-1}$. In this configuration, a cold beryllium filter was installed in the path of the incident beam to eliminate the contamination from higher-order neutrons. As a consequence, the accessible energy-transfer range was restricted to 0.9\,meV by the Be-filter cutoff energy.

The rest of the measurements were done in the conventional triple-axis configuration with a PG analyzer and the Be filter installed on $k_{\rm f}$. The same fixed value of $k_{\rm f}=1.40$\,\AA$^{-1}$ was selected. The data were corrected for the energy-dependent fraction of higher-order neutrons on the monitor. The imaginary part of the dynamical spin susceptibility, $\chi''(\mathbf{Q},\omega)$, was obtained from the background-subtracted INS intensity by the fluctuation-dissipation relation $\chi''(\mathbf{Q},\omega)=(1-\mathrm{e}^{-\hslash\omega/k_\textup{B}T})\,S(\mathbf{Q},\omega)$, where $S(\mathbf{Q},\omega)$ is the scattering function. The error bars in all figures correspond to one standard deviation of the count rate.

Throughout the paper, we index the reciprocal-lattice vectors on the simple-cubic unit cell (space group $Pm3m$, lattice constant $a=4.1367$\,\AA) and follow the conventional notation in labeling the high-symmetry points (Fig.\,\ref{Fig:RawData}e, inset). The wave-vector coordinates are given in reciprocal lattice units (1~r.l.u.~=~$2\piup/a$).

~\vspace{-0.5em}

\noindent\textbf{\sffamily Acknowledgements.} We are grateful to S.~V.~Demishev, D.~Efremov, V.~Hinkov, G.~Jackeli, G.~Khaliullin, J.-M. Mignot, Y.~Sidis, O.~Sushkov and P.~Thalmeier for stimulating discussions, valuable suggestions and encouragement. We also thank A.~Hiess, M.~Ohl, P.~Steffens and E.~Villard for technical support during the experiment and sample preparation.\smallskip

\noindent{\textbf{\sffamily Author~contributions.} N.\,E.\,S. suggested the initial idea. A.V.\,D., N.\,Yu.\,S. and V.~B.\,F. prepared the samples. G.\,F., Y.\,L., A.\,I. and D.\,S.\,I. performed the INS experiments and analyzed the data. G.\,F., Y.\,L., N.\,E.\,S., B.\,K. and D.\,S.\,I. developed the physical interpretation. A.\,I. provided instrument support at ILL. G.\,F., B\,K., and D.\,S.\,I. designed the figures and wrote the manuscript. B.\,K. and D.\,S.\,I. supervised the project.}\smallskip

\noindent{\textbf{\sffamily Author information.} The authors declare no competing financial interests. Correspondence and requests for materials should be addressed to D.\,S.\,I. $\langle$\href{mailto:d.inosov@fkf.mpg.de}{d.inosov@fkf.mpg.de}$\rangle$.}\bigskip\vfill

\vspace{1em}
\vfill
}

\onecolumngrid\clearpage

\vfill
\clearpage

\renewcommand\thefigure{S\arabic{figure}}
\renewcommand\thetable{S\arabic{table}}
\renewcommand\theequation{S\arabic{equation}}
\renewcommand\bibsection{\section*{\sffamily\bfseries\footnotesize Supplementary References\vspace{-6pt}\hfill~}}

\citestyle{supplement}

\pagestyle{plain}
\makeatletter
\renewcommand{\@oddfoot}{\hfill\bf\scriptsize\textsf{S\thepage}}
\renewcommand{\@evenfoot}{\bf\scriptsize\textsf{S\thepage}\hfill}
\renewcommand{\@oddhead}{G.~Friemel \textit{et~al.}\hfill\Large\textsf{\textcolor{NatureBlue}{SUPPLEMENTARY INFORMATION}}}
\renewcommand{\@evenhead}{G.~Friemel \textit{et~al.}\Large\textsf{\textcolor{NatureBlue}{SUPPLEMENTARY INFORMATION}}\hfill}
\makeatother
\setcounter{page}{1}\setcounter{figure}{0}\setcounter{table}{0}\setcounter{equation}{0}

\makeatletter\immediate\write\@auxout{\string\bibstyle{my-apsrev}}\makeatother

\onecolumngrid\normalsize

\begin{center}{\vspace{-3em}\large{Supplemental materials to the paper\\\sl\textbf{``\hspace{1pt}Resonant magnetic exciton mode in the heavy-fermion antiferromagnet CeB$_6$\hspace{1pt}''}}
}\end{center}

\begin{figure*}[h]
\includegraphics[width=0.85\textwidth]{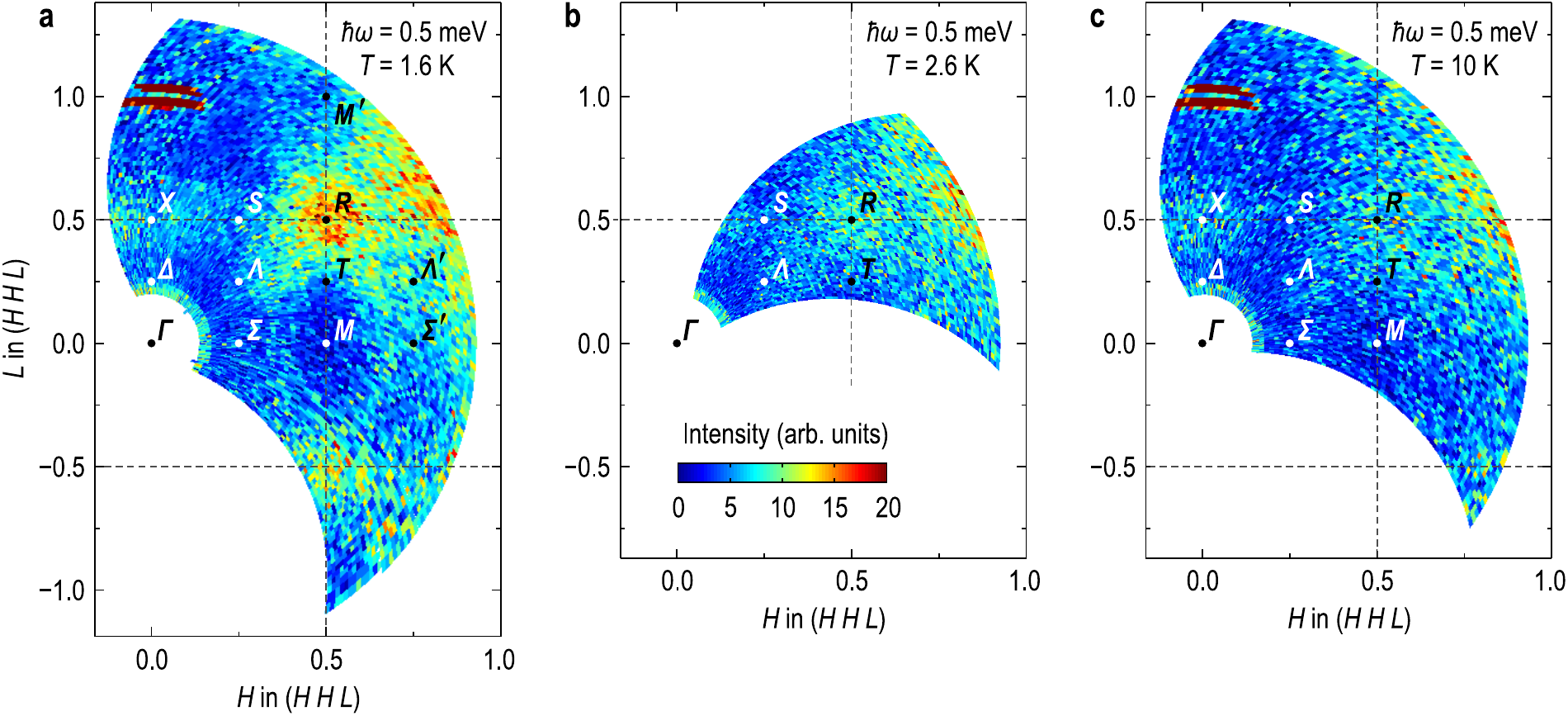}
\caption{\textbf{Temperature dependence of the inelastic intensity at $\hslash\omega=\mathbf{0.5}$\,meV.} The exciton peak at the $R(\half \half \half)$ point is significantly suppressed at $T=2.6$\,K~>\,$T_{\rm N}$ and nearly vanishes upon further warming into the normal state ($T=10$\,K), in agreement with Fig.\,\ref{Fig:Tdep}. All panels share the same color scale.\vspace{-1em}}
\label{Fig:S1}
\end{figure*}\enlargethispage{1em}

Here we present additional INS data to substantiate the temperature dependence of the new exciton mode and its shape in the reciprocal space. Additional maps presented in Fig.\,\ref{Fig:S1} above demonstrate that the strong intensity at the $R$ point (panel a) is sharply suppressed above $T_{\rm N}$ (panel b), where it transforms into the quasielastic response that continues to diminish with further temperature increase (panel c).

In Fig.~\ref{Fig:S2}a, we directly compare energy scans (raw INS intensity) at the $X$ and $\Lambda$ points, measured in the low-temperature AFM state ($T=1.6$\,K) and in the AFQ phase above $T_{\rm N}$ ($T=2.6$\,K). The low-temperature datasets are identical to those shown in Fig.\,\ref{Fig:RawData}e. At both points, the transfer of the low-energy spectral weight into a pile-up peak above the spin gap can be observed. At the $\Lambda$ point, this peak is centered at $\sim$\,0.5\,meV and represents the ``tail'' of the resonant excitation centered at $R$, which demonstrates the absence of any notable dispersion of the exciton mode.

\begin{SCfigure*}[][b]
\raisebox{20.2em}{\bf a}\hspace{-0.8em}\raisebox{0.45em}{\includegraphics[height=20.4em]{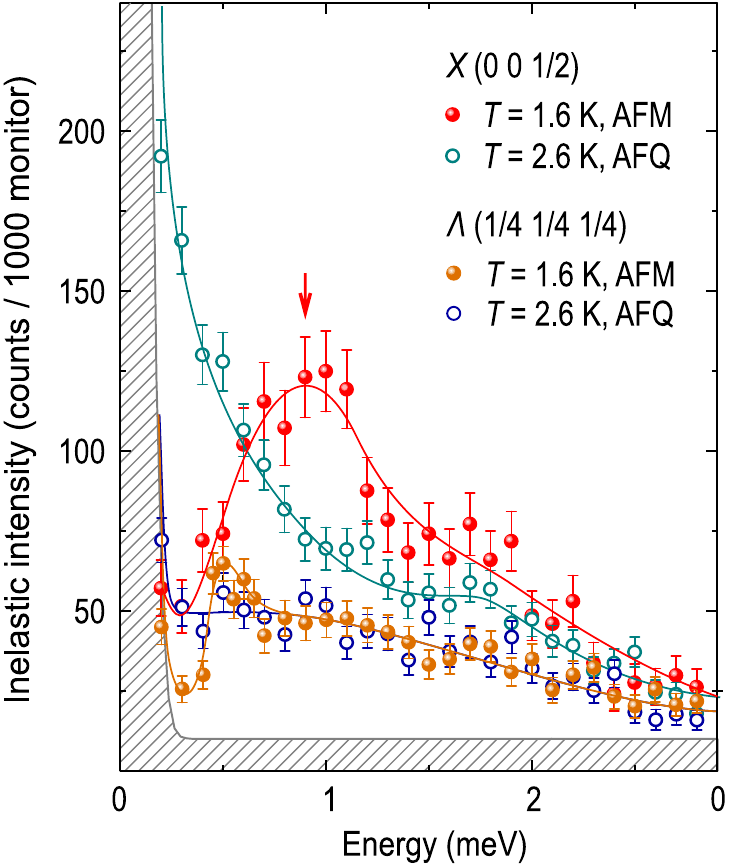}}\quad
\raisebox{20.2em}{\bf b}\hspace{-0.8em}\includegraphics[height=21em]{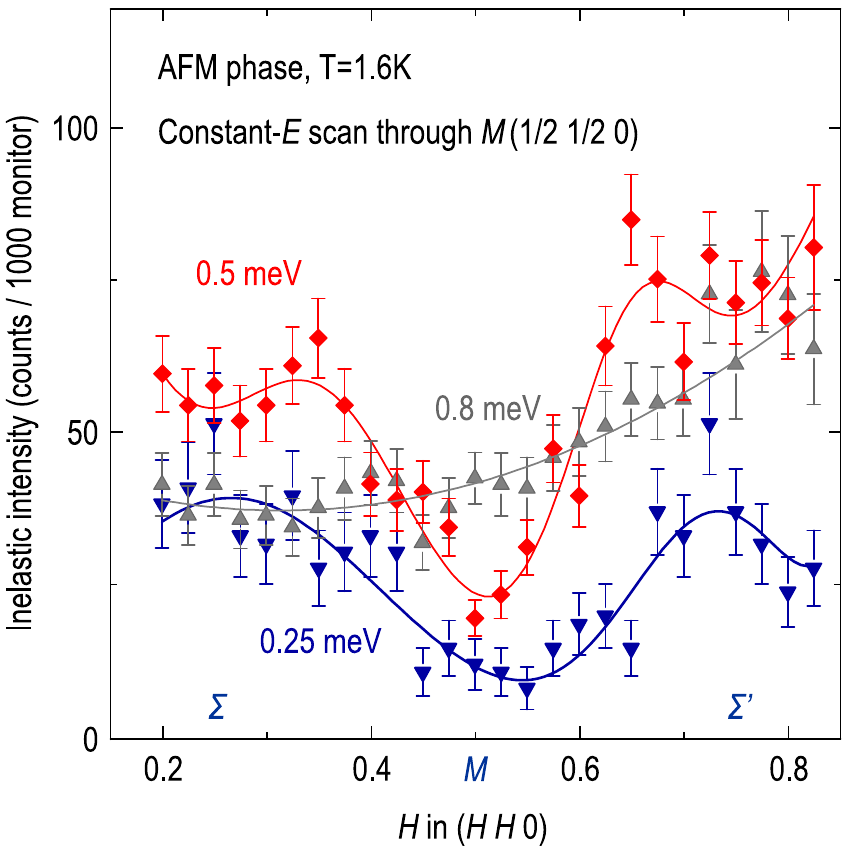}~\quad
\caption{\textbf{a},~Comparison of the energy scans measured at the $X$ and $\Lambda$ points above and below $T_{\rm N}$. \textbf{b},~Comparison of low-temperature momentum scans along the $(HH0)$ direction, centered at the $M$ point, for various energies. The scan at 0.5\,meV crosses the two intensity lobes that connect the resonant peaks at equivalent $R$ points.\vspace{1em}}
\label{Fig:S2}
\end{SCfigure*}

In Fig.~\ref{Fig:S2}b, we show additional constant-energy scans along the $(HH0)$ direction, centered at the $M$ point. Despite the fact that this scan is shifted from the $R$ point by half the Brillouin zone size, the peak intensity is still maximized at 0.5\,meV, with a deep minimum at the $M$ point. This indicates that the two nearly non-dispersing lobes of intensity that extend from the exciton peak (see Fig.\,\ref{Fig:RawData}a and Fig.\,\ref{Fig:QScans}) close into an elongated donut-shaped structure in the $(H,L,\omega)$ space, which surrounds the $M$ point and is restricted to a very narrow energy range. At a lower energy of 0.25\,meV, the intensity is partially suppressed due to the presence of the spin gap, whereas at higher energies the $\mathbf{Q}$-dependence of the intensity rapidly vanishes, as illustrated here by the 0.8\,meV data.


\smallskip

\begin{thebibliography}{10}\footnotesize
\bibitem{Eschrig06} Eschrig, M. The effect of collective spin-1 excitations on electronic spectra in high-$T_{\rm c}$ superconductors.
    \href{http://www.tandfonline.com/doi/abs/10.1080/00018730600645636}{\emph{Adv. Phys.} \textbf{55}, 47--183 (2006)}.
\bibitem{HinkovBourges07} Hinkov, V. \textit{et al.} Spin dynamics in the pseudogap state of a high-temperature superconductor.
    \href{http://www.nature.com/nphys/journal/v3/n11/abs/nphys720.html}{\textit{Nature Phys.} \textbf{3}, 780--785 (2007)}.
\bibitem{InosovPark10} Inosov, D.~S. \emph{et~al.} Normal-state spin dynamics and tempe\-rature-dependent spin-resonance energy in optimally doped {BaFe$_{1.85}$Co$_{0.15}$As$_2$}. \href{http://www.nature.com/nphys/journal/v6/n3/abs/nphys1483.html}{\emph{Nature Phys.} \textbf{6}, 178--181 (2010)}.
\bibitem{StockBroholm08} Stock, C. \emph{et~al.} Spin resonance in the $d$-wave superconductor CeCoIn$_5$.
    \href{http://link.aps.org/abstract/PRL/v100/e087001}{\emph{Phys. Rev. Lett.} \textbf{100}, 087001 (2008)}.
\bibitem{SatoAso01} Sato, N.\,K. \emph{et~al.} Strong coupling between local moments and superconducting ``heavy'' electrons in UPd$_2$Al$_3$.
    \href{http://www.nature.com/nature/journal/v410/n6826/abs/410340a0.html}{\emph{Nature} \textbf{410}, 340--343 (2001)}.
\bibitem{StockertArndt10} Stockert, O. \emph{et~al.} Magnetically driven superconductivity in CeCu$_2$Si$_2$.
    \href{http://www.nature.com/nphys/journal/v7/n2/abs/nphys1852.html}{\emph{Nature Phys.} \textbf{7}, 119--124 (2010)}.
\bibitem{MaierGraser09} Maier, T.\,A., Graser, S., Scalapino, D.\,J., Hirschfeld, P. Neutron scattering resonance and the Fe-pnictide superconducting gap.
    \href{http://link.aps.org/abstract/PRB/v79/e134520}{\emph{Phys. Rev.~B} \textbf{79}, 134520 (2009)}.
\bibitem{OnariKontani10} Onari, S., Kontani, H. \& Sato, M. Structure of neutron-scattering peaks in both $s_{++}$-wave and $s_{\pm}$-wave states of an iron pnictide superconductor, \href{http://link.aps.org/abstract/PRB/v81/e060504}{\emph{Phys. Rev.~B} \textbf{81}, 060504(R) (2010).}
\bibitem{Chubukov08} Chubukov, A.\,V. \& Gorkov, L.\,P. Spin resonance in three-dimensional superconductors: The case of CeCoIn$_5$.
    \href{http://link.aps.org/abstract/PRL/v101/e147004}{\emph{Phys. Rev. Lett.} \textbf{101}, 147004 (2008)}.
\bibitem{Eremin08} Eremin, I., Zwicknagl, G., Thalmeier, P. \& Fulde, P. Feedback spin resonance in superconducting CeCu$_2$Si$_2$ and CeCoIn$_5$.
    \href{http://link.aps.org/abstract/PRL/v101/e187001}{\emph{Phys. Rev. Lett.} \textbf{101}, 187001 (2008)}.
\bibitem{Effantin85} Effantin, J., Rossat-Mignod, J., Burlet, P., Bartholin, H., Kunii, S. \& Kasuya T. Magnetic phase diagram of CeB$_6$.
    \href{http://dx.doi.org/10.1016/0304-8853(85)90382-8}{\emph{J. Magn. Magn. Mater.} \textbf{47--48}, 145--148 (1985)}.
\bibitem{GoodrichYoung04} Goodrich, R.\,G. \emph{et~al.} Extension of the temperature-magnetic field phase diagram of CeB$_6$.
    \href{http://link.aps.org/abstract/PRB/v69/e054415}{\emph{Phys. Rev.~B} \textbf{69}, 054415 (2004)}.
\bibitem{HallFisk00} Hall, D., Fisk, Z. \& Goodrich, R. G. Magnetic-field dependence of the paramagnetic to the high-temperature magnetically ordered phase transition in CeB$_6$. \href{http://link.aps.org/abstract/PRB/v62/e84}{\emph{Phys. Rev.~B} \textbf{62}, 84--86 (2000)}.
\bibitem{NakaoMagishi01} Nakao, H. \emph{et~al.} Antiferro-quadrupole ordering of CeB$_6$ studied by resonant X-ray scattering.
    \href{http://jpsj.ipap.jp/link?JPSJ/70/1857/}{\emph{J.~Phys. Soc. Jpn.} \textbf{70}, 1857--1860 (2001)}.
\bibitem{TanakaStaub04} Tanaka, Y. \emph{et~al.} Direct and quantitative determination of the orbital ordering in CeB$_6$ by X-ray diffraction.
    \href{http://iopscience.iop.org/0295-5075/68/5/671}{\emph{Europhys. Lett.} \textbf{68}, 671--677 (2004)}.
\bibitem{MatsumuraYonemura09} Matsumura, T., Yonemura, T., Kunimori, K. Sera, M. \& Iga, F. Magnetic field induced 4$f$ octupole in CeB$_6$ probed by resonant X-ray diffraction.
    \href{http://link.aps.org/abstract/PRL/v103/e017203}{\emph{Phys. Rev. Lett.} \textbf{103}, 017203 (2009)}.
\bibitem{WooDai06} Woo, H. \emph{et~al.} Magnetic energy change available to superconducting condensation in optimally doped YBa$_2$Cu$_3$O$_{6.95}$.
    \href{http://www.nature.com/nphys/journal/v2/n9/abs/nphys394.html}{\emph{Nature Phys.} \textbf{2}, 600--604 (2006)}.
\bibitem{GrushkoPaderno85} Grushko, Yu.\,S. \textit{et al.} A study of the electronic structure of rare earth hexaborides.
    \href{http://onlinelibrary.wiley.com/doi/10.1002/pssb.2221280225/abstract}{\emph{Phys. Stat. Sol. (b)} \textbf{128}, 591--597 (1985)}.
\bibitem{PlakhtyRegnault05} Plakhty, V.~P. \emph{et~al.} Itinerant magnetism in the Kondo crystal CeB$_6$ as indicated by polarized neutron scattering.
    \href{http://link.aps.org/abstract/PRB/v71/e100407}{\emph{Phys. Rev.~B} \textbf{71}, 100407 (2005)}.
\bibitem{ZaharkoFischer03} Zaharko, O. \emph{et~al.} Zero-field magnetic structure in CeB$_6$ reinvestigated by neutron diffraction and muon spin relaxation.
    \href{http://link.aps.org/abstract/PRB/v68/e214401}{\emph{Phys. Rev.~B} \textbf{68}, 214401 (2003)}.
\bibitem{ShiinaShiba97} Shiina, R., Shiba, H. \& Thalmeier, P. Magnetic-field effects on quadrupolar ordering in a $\Gamma_8$-quartet system CeB$_6$. \href{http://jpsj.ipap.jp/cgi-bin/getarticle?journal=JPSJ&volume=66&page=1741}{\emph{J.~Phys.~Soc.~Jpn.} \textbf{66}, 1741--1755 (1997)}.
\bibitem{ThalmeierShiina03} Thalmeier, P. \emph{et~al.} Temperature and field dependence of multipolar excitations in CeB$_6$. \href{http://jpsj.ipap.jp/cgi-bin/getarticle?journal=JPSJ&volume=72&page=3219}{\emph{J.~Phys.~Soc.~Jpn.} \textbf{72}, 3219--3225 (2003)}.
\bibitem{SluchankoBogach07} Sluchanko, N.~E. \emph{et~al.} Enhancement of band magnetism and features of the magnetically ordered state in the CeB$_6$ compound with strong electron correlations.
    \href{http://www.springerlink.com/content/1725727448q8rh34/}{\emph{J.~Exp. Theor. Phys.} \textbf{104}, 120--138 (2007)}.
\bibitem{SluchankoBogach08} Sluchanko, N.~E. \emph{et~al.} Magnetization anisotropy in the AFM and SDW phases of CeB$_6$.
    \href{http://www.springerlink.com/content/b61g56j32631wg37/}{\emph{JETP Lett.} \textbf{88}, 318--321 (2008)}.
\bibitem{DemishevSemeno09} Demishev, S.~V. \emph{et~al.} Magnetic spin resonance in CeB$_6$.
    \href{http://link.aps.org/abstract/PRB/v80/e245106}{\emph{Phys. Rev.~B} \textbf{80}, 245106 (2009)}.
\bibitem{AnisimovBogach09} Anisimov, M.~A. \emph{et~al.} Magnetoresistance and magnetic ordering in praseodymium and neodymium hexaborides.
    \href{http://www.springerlink.com/content/v66l4310641q48r4/}{\emph{JETP} \textbf{109}, 815--832 (2009)}.
\bibitem{HornSteglich81} S. Horn \emph{et~al.} The magnetic behavior of CeB$_6$: Comparison between elastic and inelastic neutron scattering, initial susceptibility and high-field magnetization. \href{http://www.springerlink.com/content/g6950l6701712088/}{\emph{Z. Phys. B\,---\,Cond. Matter} \textbf{42}, 125--134 (1981)}.
\bibitem{BouvetThesis} Bouvet, A. \textit{\'{E}tude par diffusion in\'{e}lastique de neutrons des propri\'{e}t\'{e}s magn\'{e}tiques de borures de terre rare: CeB$_6$, PrB$_6$ et YbB$_{12}$.}
    PhD Thesis, Universit\'{e} de Grenoble (1993).
\bibitem{RegnaultErkelens88} Regnault, L. P. \emph{et~al.} Inelastic neutron scattering study of the rare earth hexaboride CeB$_6$.
    \href{http://dx.doi.org/10.1016/0304-8853(88)90439-8}{\emph{J.~Magn. Magn. Mater.} \textbf{76}\,\&\,\textbf{77}, 413--314 (1988)}.
\bibitem{RobertMignot10} Robert, J. \emph{et~al.} Long-range order and low-energy magnetic excitations in CeRu$_2$Al$_{10}$ studied via neutron scattering.
    \href{http://link.aps.org/abstract/PRB/v82/e100404}{\emph{Phys. Rev.~B} \textbf{82}, 100404 (2010)}.
\bibitem{Kunii87} Kunii, S. Point-contact spectroscopy of mutual \textit{RE}\hspace{0.5pt}B$_6$ (\textit{RE} = La, Y, Sm, Ce) by automatic in-situ cleaning.
    \href{http://dx.doi.org/10.1016/0304-8853(87)90700-1}{\emph{J.~Magn. Magn. Mater.} \textbf{63}, 673 (1987)}.
\bibitem{LeMcEwen08} Le, M.~D., McEwen, K.~A., Park, J.-G., Lee, S., Iga, F. \& Rule, K.~C. Magnetic excitations in the ordered phases of praseodymium hexaboride. \href{http://iopscience.iop.org/0953-8984/20/10/104231}{\emph{J.~Phys.: Condens. Matter} \textbf{20}, 104231 (2008)}.
\bibitem{CeIn3} Knafo, W., Raymond, S., F{\aa}k, B., Lapertot, G., Canfield, P.\,C. \& Flouquet, J. Study of low-energy magnetic excitations in single-crystalline CeIn$_3$ by inelastic neutron scattering. \href{http://iopscience.iop.org/0953-8984/15/22/308}{\textit{J. Phys.: Condens. Matter} \textbf{15}, 3741--3749 (2003)}.
\bibitem{CePd2Si2} van Dijk, N. H., F{\aa}k,  B., Charvolin, T., Lejay, P. \& Mignot, J.\,M.
    \href{http://link.aps.org/abstract/PRB/v61/e8922}{\emph{Phys. Rev.~B} \textbf{61}, 8922--8931 (2000)}.
\bibitem{LoewenhauptCarpenter85} Loewenhaupt, M., Carpenter, J.\,M. \& Loong, C.-K. Magnetic excitations in CeB$_6$.
    \href{http://dx.doi.org/10.1016/0304-8853(85)90270-7}{\emph{J.~Magn. Magn. Mater.} \textbf{52}, 245--249 (1985)}.
\bibitem{AlekseevMignot95} Alekseev, P.~A. \emph{et~al.} Magnetic excitation spectrum of mixed-valence SmB$_6$ studied by neutron scattering on a single crystal.
    \href{http://iopscience.iop.org/0953-8984/7/2/007}{\emph{J.~Phys.: Cond. Matter} \textbf{7}, 289 (1995)}.
\bibitem{MignotAlekseev05} Mignot, J.-M. \emph{et~al.} Evidence for short-range antiferromagnetic fluctuations in Kondo-insulating YbB$_{12}$.
    \href{http://link.aps.org/abstract/PRL/v94/e247204}{\emph{Phys. Rev. Lett.} \textbf{94}, 247204 (2005)}.
\end{thebibliography}
\end{document}